\documentclass[11pt]{article}
\usepackage{amsmath}
\usepackage{amsfonts}
\usepackage{amssymb}
\usepackage{epsfig}
\usepackage{times}
\newcommand{\be}{\begin{equation}}
\newcommand{\ee}{\end{equation}}
\newcommand{\ba}{\begin{array}}
\newcommand{\ea}{\end{array}}
\newcommand{\bea}{\begin{eqnarray}}
\newcommand{\eea}{\end{eqnarray}}

\def\Mgg{\widetilde{M}_G}
\def\Mg2{\widetilde{M}_G^2}

\def\BRA{\left\langle }
\def\KET{\right\rangle}
\def\g{\rm g_s}
\def\ket#1{\,\left\vert  #1\right\rangle}
\def\bra#1{\left\langle#1\right\vert\, }

\def\BO{\BRA{0 \vert} } 
\def\KO{\vert 0 \KET   } 

\def\b{\rm b}
\def\a{\rm a}
\def\D{\rm D}
\def\c{\rm c}
\def\d{\rm d}
\def\dpsi{[\D \psi]}
\def\Dpsi{[\D \overline{\psi}]}
\def\dchi{[\D \chi]}
\def\Dchi{[\D \overline{\chi}]}
\def\dPhi{[\D \Phi]}
\def\DPhi{[\D \overline{\Phi}]}
\def\e{\rm e}
\def\E{\rm exp}

\def\hbar{\not{\hbox{\kern-2.3pt $h$}}}
\def\psl{\not{\hbox{\kern-2.3pt $p$}}}
\def\Psl{\not{\hbox{\kern-2.3pt $P$}}}
\def\ksl{\not{\hbox{\kern-2.3pt $k$}}}
\def\qsl{\not{\hbox{\kern-2.3pt $q$}}}
\def\slad{\not{\hbox{\kern-2.3pt $\partial$}}}
\def\I{\rm i}
\begin{document}
\begin{titlepage}

January 2001 \hfill PAR-LPTHE 01/03
\vskip 4.5cm
{\baselineskip 17pt
\begin{center}
{\bf GLUON-PHOTON MIXING in DENSE QCD}
\end{center}
}

\vskip 1cm
\centerline{
Nguyen Van Hieu 
     \footnote[1]{E-mail: nvhieu@ims.ncst.ac.vn}
}
\vskip 2mm
{\em 
Institute of Physics, NCST of Vietnam}  
     \footnote[2]{ Nghia Do, 
Hoang Quoc Viet Str., Cau Giay, Hanoi 
(Vietnam)}
{\em and  Faculty of Technology, Hanoi National University } 
\footnote[3]{  144 Xuan Thuy Str., Cau Giay, Hanoi (Vietnam)}
\vskip3mm
\vskip2mm
\centerline{\em  and}
\vskip3mm
\centerline{
Pham Xuan-Yem
     \footnote[4]{E-mail: pham@lpthe.jussieu.fr}
            }
\vskip 2mm
{\em Laboratoire de Physique Th\'eorique et Hautes Energies, 
Unit\'e associ\'ee au CNRS, 
 }
     \footnote[5]{LPTHE tour 16\,/\,1$^{er}\!$ \'etage,
         4 place Jussieu,
          F-75252 Paris Cedex 05 (France)}

\centerline{\em Universit\'es Paris 6 et Paris 7} 
\vskip 1.5cm

{\bf Abstract:}.  At 
high baryonic density with the formation of a diquark condensate 
$\Delta \neq 0$, the 
QCD color symmetry is spontaneously broken. Being massive by the 
Anderson-Higgs mechanism, gluon and photon should mix 
together within two linear combinations due to 
the color nonconservation. Consequently a gluon 
$\widetilde{G}$ could decay into 
an $e^- e^+$ pair via its photon component.

 With a low invariant mass (about a few ten MeV) and an 
extremely narrow width peaking above the  continuum 
background, the purely leptonic decay of a strongly-interacting gluon 
$\widetilde{G} \to e^- + e^+$  constitutes 
a very distinctive signature of the 
color superconductivity phase. By a similar scenario of 
gluon-$Z$ mixing, another  
 "missing-energy"  decay into invisible neutrinos
 $\widetilde{G} \to \nu +\overline{\nu}$
could arise, its amplitude is however  $(\Delta/M_Z)^2$ 
power-suppressed.

\smallskip

{\bf PACS} number(s): 24.85.+p , \quad 11.15.Ex , \quad 12.38. Aw 
\end{titlepage}
%
%
\section{Introduction}

 In these last years, our understanding of QCD at extreme conditions, 
in particular at high baryon density and low temperature, has been 
considerably advanced and these extensive 
developments\cite{alford} -- \cite{hieu} are amply discussed in some recent 
reviews\cite{raja}.

 Our starting point is a simple observation that at sufficiently 
high density characterized by a chemical potential $\mu \neq 0$, 
quark matter is a color superconductor\cite{alford}--
\cite{Barrois} similar to 
the standard Bardeen-Cooper-Schrieffer (BCS) superconductivity 
of metals. Thus like 
a Cooper electron pair, the  pairing of up and down light quarks 
either by gluon exchange or by instanton-induced would lead to a 
diquark condensation $\Delta \neq 0$ which breaks the 
color symmetry $SU_c(3)$ down to an unbroken $SU_c(2)$ 
color subgroup. 

We are in the so called two flavors superconductivity
${\bf 2SC}$ phase with two flavors ($i, j = 1,2$ 
stand for up and down quarks) linked by their 
three colors ($\alpha, \beta, \gamma  = 1,2,3$) to form 
a quark bilinear combination $ \Phi_{\gamma} = 
\varepsilon^{i j} \;\varepsilon_{\alpha
 \beta \gamma}  q_i^\alpha q_j^\beta$  which behaves as 
a Lorentz  scalar-isoscalar complex field  belonging to a 
fundamental representation  
 of the $SU_c(3)$ group as an $\overline{3}$ color antitriplet. 
Its nonzero vacuum expectation value designating the diquark condensate 
takes the form 
\begin{equation}
 \BRA q_i^\alpha q_j^\beta \KET \varepsilon^{i j} 
\;\varepsilon_{\alpha \beta \gamma} =
\Delta_{\gamma} 
\;.
\end{equation}

QCD gauge invariance implies that the composite $\Phi_\gamma$  field 
couples to the eight gluons  $ G^a$ via the covariant derivative
$(\bigtriangledown_\rho)_\alpha^\beta =\partial_\rho
 \delta_\alpha^\beta - {\rm i} \g  G^a_\rho 
(\lambda^a/2)_\alpha^\beta$ in which $\lambda^a$ denote
the Gell-Mann matrices of $SU_c(3)$.
Since color is spontaneously broken by the 
order parameter $\Delta \neq 0$, five
among the eight $G^a$ gluons  
become massive via the Anderson-Higgs-Meissner mechanism, 
the other three gluons belonging to the unbroken $SU_c(2)$ 
color subgroup remain massless, as it should be. 

Furthermore when electromagnetism is switched on, the covariant 
derivative $\bigtriangledown_\rho$ has an additional 
$-\I \e A_\rho (\tau^c)$ term, $\tau^c$ is the charge flavor 
matrix defined later in (7). Due to the color symmetry breaking, 
the colorfull gluons $G^a$ and the colorless photon $A$ 
are necessarily mixed together. The mixing asises from the 
diquark condensate $\Delta$ inserted in all possible ways as 
depicted in Figs.I. This case cannot 
happen to  QCD at zero density since $\Delta$ 
being proportional to the chemical potential $\mu$ like 
$\Delta \sim \mu \;\E({-3\pi^2/\sqrt{2} \g}) $ 
according to\cite{son1},\cite{pis} and tends to zero as $\mu \to 0$. 
The gluon-photon mixing allows us to define 
a "physical" mass-eigenstate gluon 
$\widetilde{G^a},\; a = 4,...,8$ as a linear combination of 
two gauge-eigenstates 
with necessarily the photon $A$ as one inevitable 
component. Thus 
\begin{equation}
 \widetilde{G^a} = \cos \theta_a \, G^a \; - \,\sin \theta_a \, A
\;,
\end{equation}
 the mixing angle $\theta_a$ is function of the condensate $\Delta$, 
 the strong QCD coupling constant $\g$ and the quark electromagnetic
 coupling constants $\e Q_j$ (Q$_u =2/3$ and Q$_d = -1/3)$ as will follow 
below. The orthogonal combination
\begin{equation}
 \widetilde{A} = \cos \theta_a \, A\; + \,\sin \theta_a \, G^a
\;
\end{equation}
is the "rotated" physical photon in the ${\bf 2SC}$ phase.
The
 $a=8$ gluon case associated with the diagonalized 
$\lambda^8$ matrix has been considered in\cite{alford2}.
 
The situation is very similar to the standard electroweak interactions 
in which the massive $Z$ weak neutral boson is a linear combination 
 of two gauge boson eigenstates, i.e. the  
hypercharge isosinglet $B$ is mixed with the third component $ W^3$ of 
the $SU_L(2)$ isotriplet ${\bf W}$ to build up the $Z$, thus
\begin{equation}
 Z = {g \over \sqrt{g^2 +g'{^2}}} W^3  \;\; - \;\; 
{g'\over \sqrt{g^2 +g'{^2}}} B
\;,
\end{equation}
 the ratio $g'/\sqrt{g^2 +g'{^2}}$ is defined as $ \sin \theta_W$, where 
$\theta_W$ is the Weinberg angle.
Neither the $ W^3$ nor the $B$ has a well-defined  mass, 
only their mixture $Z$ is a mass-eigenstate by the 
diagonalization procedure.

The gluon-photon mixing ($\theta_a$, $\widetilde{G}$) shares  
with the standard electroweak interactions ($\theta_W$, $Z$) 
and the BCS superconductivity a common crucial point: 
respectively in these 
three phenomena, gauge symmetries are spontaneously broken 
with nonzero vacuum expectation values $\BRA {\rm u}^\alpha 
{\rm d}^\beta - {\rm d}^\alpha 
{\rm u}^\beta \KET \neq 0$ of the up-down diquark, $v \neq 0$  
of the elementary scalar Higgs field and $\BRA e^- e^- \KET \neq 0$ 
of the Cooper electron pair.  

We also note that in the standard electroweak interactions, 
the decays of $Z$ come from those of both 
 $ W^3$  and $B$. 
For instance the $Z$ decaying into a neutrino pair is due only
to its $ W^3$ component, and the decay of $Z$ into a right-handed 
electron pair is due only to its $B$ part. 
 Similarly, a massive gluon $\widetilde{G}$ could decay 
electromagnetically into $e^+ + e^-$ via its photon component 
as shown by (2), otherwise this process is strictly forbidden in 
non dense media (chemical potential $\mu = 0$). 
 This  lepton pair production 
by copious gluons in quark matter 
may provide a remarkable signature 
 of  the  ${\bf 2SC}$ phase transition in neutron star cores  
and perhaps under certain circumstances in heavy ion collisions. 

Our next task is devoted  to an estimate of the gluon mass 
$\widetilde{M}_G$, its mixing angle  and its 
lepton pair decay width $\Gamma (\widetilde{G} \to e^+ + e^-)$. We 
will see that this lepton pair,  with a small invariant mass 
about 90 MeV and an extremely narrow width is definitely distinct from 
the continuum Drell--Yan background and the leptonic decays of the 
well-known vector mesons $\rho(770), \omega(782)$
 with their peaks and tails modified by medium effects\cite{sarkar}.

\section{Gluon, Photon Masses and their Mixing}

The traditional way to estimate the  photon and gluon  masses
 as well as their mixing in dense matter is to consider their 
self-energies via their corresponding polarization 
tensors $\Pi_{\rho\sigma}$. The calculational method 
 adopted here follows from\cite{hieu}.
The  Lagrangian for gluon and photon interacting with quarks 
may be put in the form
\begin{equation}
 {\cal L}_{\rm int} = \g G^a_\rho  {\cal J}^a_\rho + \e A_\rho J^c_\rho
 \;. 
\end{equation}
The quark currents ${\cal J}^a_\rho$ for QCD and $J^c_\rho$ 
for QED, respectively coupled to the gluon $G^a_\rho$ and the 
photon $ A_\rho$ with the coupling constants $\g$ and $\e$,
 are given by  
\begin{equation}
{\cal J}^a_\rho  =\overline{\Psi}^A (\gamma_\rho \lambda^a) \Psi_B 
\;\;,\;\;{\rm and} \;\;
J^c_\rho =\overline{\Psi}^A (\gamma_\rho \tau^c) \Psi_B \;,
\end{equation}
where $\tau^c $ is a $2 \times 2 $ charge matrice
 in the flavor space that we introduce together with another
 antisymmetric $\varepsilon$ matrix,
\begin{equation} 
\tau^c= \left(\begin{array} {c c }
 {\rm Q}_u & 0 \cr 
 0 & {\rm Q}_d \end{array}  \right)
                   \;\;,\;\; 
\varepsilon = \left(\begin{array} {c c }
 0 & 1 \cr 
 -1 & 0 \end{array}  \right) \;.
\end{equation}
The quark-quark pairing term of the effective Lagrangian 
for the quarks in the condensate  may be written as\cite{hieu}
\begin{equation}
{\cal L}_{\rm quark}  = \overline{\Delta}^{A B}  \Psi_B \Psi_A + 
\overline{\Psi}^A \overline{\Psi}^B \Delta_{B A} 
 \;,
\end{equation}
 the indices $A, B$ stand collectively for $i,j$ flavors,  
$\alpha, \beta$ colors, and ${\cal A},{\cal B}$ Dirac spinor indices 
on which operates $\gamma_5 {\cal C}$, ${\cal C}
 =\I \gamma^2 \gamma^0$ being the charge conjugation operator. 
 Thus
$A = (i, \alpha, {\cal A}), B=(j,\beta, {\cal B}) $, and
\begin{equation}
\Delta_{AB} =\left(\gamma_5 {\cal C} \right)_{{\cal A} \;{\cal B}}
 \varepsilon_{i j} \varepsilon_{\alpha \beta \gamma}
 \Delta^{\gamma}  \;\;,\;\;\overline{\Delta}^{ AB} = \gamma_0 
\Delta^{\dagger}_{ AB}
\gamma_0 \;.
\end{equation} 
As an example, let us consider in some details the gluon-photon mixing 
manifested through the correlation function
 $\Pi^{\gamma  G^a}_{\rho \sigma}$:
\begin{equation} 
\Pi^{\gamma  G^a}_{\rho \sigma}(x-y) \equiv \BO T
 \left[ {\cal J}^a_\rho (x) 
J^c_\sigma (y) \right] \KO 
=  \bra 0  T \left\{ \overline{\Psi}^A (x) 
(\gamma_\rho \lambda^a)_A^B \Psi_B (x) \; \overline{\Psi}^C (y)
 (\gamma_\sigma \tau^c)^D_C \Psi_D (y) \right.  \cr 
  \times \; \E \int \d z
 \left. \left[\overline{\Delta}^{K L} \Psi_L (z) \Psi_K (z) + 
\overline{\Psi}^K (z) 
\overline{\Psi}^L (z) \Delta_{L K} \right] \right\}   \ket 0   \;.
\end{equation}
The development of the
 exponential to  second order in $\Delta$ leads to
\begin{equation} 
\Pi^{\gamma  G^a}_{\rho \sigma}(x-y)  \approx  {1\over 2} 
\int \d z \int \d u  \bra 0 T \left\{ \overline{\Psi}^A (x) 
(\gamma_\rho \lambda^a)_A^B \Psi_B (x) \; \overline{\Psi}^C (y)
 (\gamma_\sigma \tau^c)^D_C \Psi_D (y)  \right. \cr
 \left. \times\left[\overline{\Delta}^{K L} \Psi_L (z) \Psi_K (z) + 
\overline{\Psi}^K (z) 
\overline{\Psi}^L (z) \Delta_{L K} \right] \times
\left[\overline{\Delta}^{M N} \Psi_N (u) \Psi_M (u) + 
\overline{\Psi}^M (u) 
\overline{\Psi}^N (u) \Delta_{N M} \right] \right\}  \ket 0 \;,
\end{equation}
which can be rewritten as
\begin{equation} 
(\gamma_\rho \lambda^a)_A^B \; (\gamma_\sigma \tau^c)^D_C \;
\overline{\Delta}^{K L}\,\Delta_{N M} \cr
\times \int \d z \int \d u  \bra 0 T \left\{ \overline{\Psi}^A (x) 
 \Psi_B (x) \; \overline{\Psi}^C (y)
\Psi_D (y) \; \Psi_L (z) \Psi_K (z) \overline{\Psi}^M (u) 
\overline{\Psi}^N (u)\right\} 
 \ket 0 \;.
\end{equation}
When the Wick's theorem for the time-order products is applied  
to  the above equation and the Fourier transformation in momentum 
space is made, we get three following contributions written 
in  a rather compact form and  
depicted respectively by the three diagrams of Figs.I. They are
\begin{equation} 
\Sigma_1 \times (\lambda^a)_A^B \Delta_{B C}\;
\overline{\Delta}^{C D}\;(\tau^c)^A_D  
\;\;,\;\; \Sigma_2 \times (\lambda^a)_A^B \;(\tau^c)^C_B \;
 {\Delta}_{C D}\;
\overline{\Delta}^{DA} \; \;,\;\; \Sigma_3 \times (\lambda^a)_A^B\;
 {\Delta}_{B C} 
\;(\tau^c)^C_D \;\overline{\Delta}^{DA} \;,
\end{equation}
the coefficients $\Sigma_{n}$ with $n = 1,2,3$
 are obtained from the three 
corresponding quark loop integrations, an overall factor 
$( \e \g)$ must be included.
After some algebraic manipulations, we get 
\begin{equation} 
\Pi^{\gamma G^a}_{\rho \sigma}(p)  =  
 - \e\,\g {\rm Tr} (\tau^c)\left( \Delta^\alpha 
\left(\lambda^a \right)^\beta_\alpha \overline{\Delta}_\beta \right)
\left[ U^{(1)}_{\rho \sigma}(p) + U^{(2)}_{\rho \sigma}(p) + 
U^{(3)}_{\rho \sigma}(p) \right]\;,
\end{equation}
where
\begin{equation} 
 U^{(1)}_{\rho \sigma}(p) ={\I \over (2\pi)^4} \int \d^4 k\; {\rm Tr} 
\left[ \gamma_\rho \,S(p+k) \, S(-p-k)\, S(p+k) \gamma_\sigma \,S(k)
 \right]\;,\cr
\hskip-12mm U^{(2)}_{\rho \sigma}(p) ={\I \over (2\pi)^4} \int 
\d^4 k\; {\rm Tr} 
\left[ \gamma_\rho \,S(p+k) \,\gamma_\sigma S(k)\, S(-k)  \,S(k)
 \right]\;,\cr
U^{(3)}_{\rho \sigma}(p) ={\I \over (2\pi)^4} \int \d^4 k \;{\rm Tr} 
\left[ \gamma_\rho \,S(p+k) \, S(-p-k) \gamma_\sigma S(-k)\, S(k) 
 \right]\;,
\end{equation}
and $S(k)$ in (15) being the quark propagator.
Similarly we obtain for the photon-photon and  
gluon-gluon self-energy tensors the following expressions:
\begin{equation} 
\Pi^{\gamma \gamma}_{\rho \sigma}(p)  =   2 \e^2 \left\{  {\rm Tr}
 \left[(\tau^c)^2\right]
\left( \Delta^\alpha \overline{\Delta}_\alpha \right) \;
\left[ U^{(1)}_{\rho \sigma}(p) + U^{(2)}_{\rho \sigma}(p)\right]
-\, {\rm Tr}\left(\tau^c \varepsilon \tau^c \varepsilon \right) 
\left[ \Delta^\alpha \overline{\Delta}_\alpha \right]  U^{(3)}_{\rho
 \sigma}(p) \right\}
\;,
\end{equation}
and
\begin{equation} 
\hskip-28mm \Pi^{G^a G^b}_{\rho \sigma}(p)  =   2 \g^2 \left\{ {\rm Tr}
 \left[\lambda^a \lambda^b \right]
\left( \Delta^\alpha \overline{\Delta}_\alpha \right) \;
\left[ U^{(1)}_{\rho \sigma}(p) + U^{(2)}_{\rho \sigma}(p) -  
U^{(3)}_{\rho \sigma}(p) \right] \right. \cr
 \left. + \Delta^\alpha 
\left(\lambda^a \lambda^b \right)^\beta_\alpha \overline{\Delta}_\beta 
  \left[ U^{(3)}_{\rho \sigma}(p) -  
U^{(2)}_{\rho \sigma}(p) \right]
+ \Delta^\alpha 
\left(\lambda^b \lambda^a \right)^\beta_\alpha \overline{\Delta}_\beta 
  \left[ U^{(3)}_{\rho \sigma}(p) -  
U^{(1)}_{\rho \sigma}(p) \right] \right\}\;.
\end{equation}
 It remains to compute the loop integrations 
$$U^{(n)}_{\rho \sigma}(p) 
 = \left(g_{\rho \sigma} - {p_\rho p_\sigma \over
 p^2}\right) U^{(n)}(p).$$ 
We estimate  
 them at zero temperature, $\mu =0$, and get
for $p=0$ :
\begin{equation} 
 U^{(1)}_{\rho \sigma}(0) =  U^{(2)}_{\rho \sigma}(0) =
 U g_{\rho \sigma} \;\;,\;\;  U^{(3)}_{\rho \sigma}(0) =
 V g_{\rho \sigma}\;,
\end{equation}
with
\begin{equation} 
U = {1\over 2 } V =  {1\over 8 \pi^2} 
 \ln \left( {\Lambda^2 +m^2 \over m^2}
\right)\;, 
\end{equation}
where $\Lambda$ is the momentum cut-off and m being the 
effective quark mass.
The three polarization tensors  $\Pi^{\gamma G^a}_{\rho \sigma}$, 
$\Pi^{\gamma \gamma}_{\rho \sigma}$ and 
$\Pi^{G^a G^b}_{\rho \sigma}$ 
obtained above in (14), (16) and (17) give the following terms 
to the  Lagrangian of the 
gauge-eingenstates $A$ and $G^a$ :
\begin{equation} 
 {\cal L}_{\rm gauge} = {1\over 2} \left[ m^2_{\gamma} (A_\rho)^2 +
m^2_{a} (G^a_\rho)^2 + f_a A_\rho  G^a_\rho \right] \;,
\end{equation}
for any gluon  $a =4...8$.
We find
\begin{equation} 
\hskip-28mm m^2_{\gamma} = 4 \e^2 |\Delta^2| \left\{ 2 {\rm Tr} 
\left[(\tau^c)^2\right] \,U - {\rm Tr} \left[ \tau^c \varepsilon
\tau^c \varepsilon \right] \,V \right\} \;, \cr
m^2_{a} = 4 \g^2 \left\{ |\Delta^2|  {\rm Tr} \left[(\lambda^a)^2 \right] 
  (2U - V) + 2 \Delta^\alpha 
\left[(\lambda^a)^2\right]^\beta_\alpha  \overline{\Delta}_\beta 
\left(V - U \right) \right\} \;, \cr
\hskip-30mm f_a = - 2 \e \g \; \Delta^\alpha 
\left[\lambda^a\right]^\beta_\alpha  \overline{\Delta}_\beta  
\; {\rm Tr} (\tau^c) \left( 2U + V \right)  
\;.
\end{equation}
Since $\Delta_{\gamma}$ can be always made real and 
arranged along the third color
 axis ($\Delta_{\gamma} = \Delta \, \delta_{\gamma}^3)$ 
without loosing any generality\cite{carter}, 
it turns out from (19), (20) and (21) that
\begin{equation} 
 m_4 = m_5 = m_6 = m_7 = \sqrt{{3\over 4}} m_8  \equiv M_{G}
\;,
\end{equation}
in agreement with\cite{carter}, so in the following only a common $ M_{G}$ is kept.

The  ${\cal L}_{\rm gauge}$ in (20) may be conveniently 
put in a diagonalized form
 with the "physical" mass-eigenstates $\widetilde{A}$ and 
$\widetilde{G}$; their respective masses are denoted by  
$\widetilde{m}_{\gamma}$ and $\widetilde{M}_G$: 
\begin{equation} 
 {\cal L}_{\rm gauge} = {1\over 2} \left[ \widetilde{m}^2_{\gamma} 
(\widetilde{A}_\rho)^2 +
\widetilde{M}^2_G (\widetilde{G}_\rho)^2 \right]  
\;.
\end{equation}
The mass-eigenstates $\widetilde{A}$ and 
$\widetilde{G}$ are linear combinations of the gauge-eigenstates 
$A$ and $G$ through
\begin{equation} 
 \widetilde{A} = A \,\cos\theta + G\, \sin\theta \;,\cr
 \widetilde{G} = -A\,\sin\theta + G\, \cos\theta
\;,
\end{equation}
where the mixing angle $\theta$ is given by
\begin{equation} 
 \sin^2 \theta ={1\over 2} \left[ 1 - {M^2_{G} - m^2_{\gamma} \over 
\sqrt{(M^2_{G} - m^2_{\gamma})^2 +f^2}} \right]
 \;.
\end{equation}

\def\mgg{ \widetilde{M}_G }
\def\mg2{ \widetilde{M}^2_G }
 
 Moreover, since $M_G^2, m_{\gamma}^2$ and $f$ are all proportional to 
$\Delta^2$, by factorization the mixing angle $\theta$ depends only 
on the ratio $\; \e/\g$ and not on $\Delta$. 
The well-defined masses of the mass-eigenstates  $\widetilde{G}$
and $\widetilde{A}$ are
\begin{equation}
\widetilde{M}_G^2 = {1\over 2} \left[ M^2_{G} + m^2_{\gamma} + 
 \sqrt{(M^2_{G} - m^2_{\gamma})^2 +f^2} \right] \;,\cr
\widetilde{m}^2_{\gamma} = {1\over 2} \left[M^2_{G} + m^2_{\gamma} - 
 \sqrt{(M^2_{G} - m^2_{\gamma})^2 +f^2} \right]
 \;.
\end{equation}
Since the parameter $f$ is definitely nonzero from (21), let 
us remark that the mixing angle $\theta$ never vanishes 
no matter how small $\widetilde{m}_{\gamma}$ is, including 
its zero value.
It turns out that in some particular case, the mixing parameter $f$ 
could happen to be $2 M_G m_{\gamma}$, 
then $\widetilde{m}_{\gamma} = 0$, the "rotated" physical photon 
$\widetilde{\gamma}$ could 
be massless\cite{alford2} even in dense matter, although its 
associated gauge-eigenstate 
 photon $A$ has  $m_{\gamma} \neq 0$. In this case,
  $\sin\theta = m_{\gamma}/ \sqrt{M_G^2 + m^2_{\gamma}} \sim e/
\sqrt{\g^2 + \e^2}$. This situation is just as in 
the standard electroweak 
interactions where neither the weak $T_3$ nor the hypercharge $Y$ 
cancels the Higgs vacuum expectation value $v$ but only does 
their linear combination
$Q =T_3 +Y/2$. The symmetry generated by $Q$ is therefore unbroken, thus 
making the photon massless and $\sin \theta_W = g'/
\sqrt{g^2 + g'{^2}}$. Similarly here\cite{alford2}, a linear 
combination of the color $T_8$  
and the electric charge $Q$ 
is left unbroken, making $\widetilde{m}_{\gamma} =0$. 
There is an one to one correspondence between the color 
superconductivity and
 the standard electroweak interactions: $(\widetilde{G}, 
\widetilde{\gamma})
 \Longleftrightarrow (Z, \gamma)$, $ (G, \gamma) 
\Longleftrightarrow (W^3, B)$,
$ (\g, \e) \Longleftrightarrow (g, g')$.

Numerical values for the masses $\widetilde{M}_G, 
\widetilde{m}_{\gamma} $  depend on $\Delta(\mu)$, 
the strong QCD coupling constant $\g$, and the $U, V$ terms, the 
latter presumably have a smooth\cite{carter} dependence on $\mu$.
Using $\Delta \approx 200$ MeV at $\mu \approx 300$ MeV and
$\alpha_{\rm s} =\g^2/4\pi \approx 0.6$, we get
$\widetilde{M}_G  \approx 90$ MeV which is 
consistent with other evaluations\cite{carter}--\cite{son2}. 
Our value of 90 MeV is 
only an order of magnitude estimate, with large uncertainties.

 We also note from (21) that ${m}_{\gamma} /M_G 
  \approx \e/\g \ll 1$, hence  (25) and (26) 
imply that both $\widetilde{m}_{\gamma}$ and 
$\sin \theta$ are small, in particular 
\begin{equation}
 \sin\theta \approx { \e \over 2 \sqrt{3}\g} \approx 0.032\;. 
\end{equation}
As given by (24), the $\e^+ +\e^-$ decay mode of  
$\widetilde{G}$ is due to its
 photon component which carries a $(-\sin\theta)$ coefficient.
Its decay width $ \Gamma(\widetilde{G} \to \e^+ +\e^-)$ is 
computed to be
\begin{equation}
\Gamma(\widetilde{G} \to \e^+ +\e^-) = 
{\e^2 \sin^2\theta \over 12 \pi}\; \Mgg \left(
 1+ {2m_e^2 \over \Mg2} \right) \sqrt{ 1- {4m_e^2 \over \Mg2
 }} \cr
\approx  {\alpha_{\rm em}^2 \over 36 \;\alpha_{\rm s}} \Mgg 
  \approx  0.27 \; {\rm KeV} \approx 4 \times 10^{17} /{\rm sec}  \;. 
\end{equation}
To estimate the corresponding B$(\widetilde{G} \to \e^+ +\e^-)$ 
branching ratio
$${\rm B}(\widetilde{G} \to \e^+ +\e^-) \equiv {\Gamma(\widetilde{G}
 \to \e^+ +\e^-) \over \Gamma(\widetilde{G}
 \to {\rm all})}\;,$$
 we assume $ \Gamma(\widetilde{G}
 \to {\rm all}) \approx  \cos^2\theta \;\Gamma(\widetilde{G} 
 \to {\rm q} +\overline {\rm q}) \approx \alpha_s \Mgg / 3$. 
Thus
B$(\widetilde{G} \to \e^+ +\e^-) 
\approx \alpha_{\rm em}^2 / 12 \alpha_s^2 \approx 10^{-5} $. 
\section{Summary and Outlook}

 As is well- known, statistical treatment of QCD indicates that 
at high densities, the quarks and gluons will become deconfined, 
leading to a new state of matter, the so-called Quark-Gluon Plasma 
(QGP).
This plasma is however ephemeral and soon hadronizes 
into mesons and baryons so the problem is 
to find out what are the QGP distinctive signatures. Candidates 
 for these signatures as amply discussed in the literature 
(see for instance\cite{raja},\cite{sarkar}--\cite{vanei})
are 
strangeness enhancement, J/$\psi$ suppression, enhancement 
in the continuum low mass 
 region 250 MeV $< M_{\ell \ell} <$ 700 MeV of the  
lepton pair invariant  mass   
$M_{\ell \ell}$ spectra, $\ell^{\pm}$ stands for both 
muon and electron. See also\cite{raja} for astrophysical signatures.

We suggest here that another signature 
of the phase transition is 
provided by an extremely narrow peak at the  $e^+ e^-$ invariant  mass 
$M_{e^+ e^-} \approx 90$ MeV.

In summary, at high density with the chemical potential $\mu \neq 0$, 
the QCD color symmetry is spontaneously broken by the formation 
of a diquark condensate. This is similar to the BCS 
superconductivity for which no matter how weak is the attractive 
interaction between electrons due to phonic vibrations, the Cooper 
electron pair condensate breaks the QED gauge symmetry, 
thus giving a Meissner mass to the  photon.
Furthermore, because of the diquark condensate $\Delta \neq 0$, the 
mixing parameter $f$ in (22) is shown to be definitely nonzero, 
therefore the gluon and photon must necessarily mix inside a 
massive gluon $\widetilde{G}$, exactly as in the standard electroweak 
interactions where the vacuum expectation value $v\neq 0$ of the 
elementary Higgs field necessarily mix the gauge-eigenstates 
 $ W^3$ and $B$ to build up the mass-eigenstate weak neutral 
$Z$ boson. 
The $B$ component of the resulting $Z$ is responsible for 
its unusual decay  into right-handed electron + left-handed positron
  $Z \to e^{-}_{\rm R} +e_{\rm L}^{+}$, as well the
$ W^3$  component of  $Z$ is responsible of the  
$Z \to \nu +\overline{\nu}$ mode. 

Similarly, the photon component of $\widetilde{G}$ implies the 
existence of the ${\it purely \; leptonic \; gluon \; decay}$.  
The gluon-photon mixing scheme which yields $\widetilde{G} 
\to \e^+ +\e^-$ seems to provide a remarkable signature for the color 
superconductivity phase transition. 
Because of the small gluon mass $\approx $ 90 MeV, only the decay 
$\e^+ +\e^-$ is available, the other mode 
$\mu^+ +\mu^-$ is absent, in contrast to the Drell--Yan 
continuum dileptons $\ell^+ +\ell^-$  spectra emanating from the usual
 sources such as quark + 
antiquark annihilation, quark +gluon Compton scattering, 
gluon+gluon fusion, supplemented by medium effects\cite{sarkar} on known 
vector mesons $\rho (770), \omega(782)$. 
  In the gluon-photon mixing mechanism, the $\e^+ +\e^-$ 
spectrum has only one sharp peak at low mass $\approx$ 90 MeV. 
Although the 
$\widetilde{G} \to \e^+ +\e^-$ branching ratio is small 
$\approx 10^{-5}$,
 however since gluons are so abundant in quark matter, 
  this remarkable leptonic gluon decay  mode 
may be hopefully observable.

 Finally we remark that with a diquark condensate, when the 
electroweak interactions are considered together with QCD, 
the $Z$ neutral boson, like the photon, should also mix with 
the gluon, hence the mode 
$\widetilde{G} \to \nu +\overline{\nu}$  could occur. Its amplitude 
is  however $\left(\Delta/M_{Z}\right)^2$
 power-suppessed, otherwise the gluonic decay 
into invisible neutrinos, manifested by a "missing-energy" reaction, 
would be another surprising signature of the quark matter phase 
transition. 

Also at sufficiently high density  for which the 
strange quark mass may be neglected ($\mu  \gg m_s$), we are in 
the so called color-flavor locking phase ${\bf CFL}$ with three massless 
flavors locked to their three colors,  the basics
 of   purely leptonic  decays of strongly-interacting massive
 gluons 
 remain unchanged.

{\bf Acknowledgment}- One of us (N.V.H.) would like to express 
his sincere thank to the National Council for Natural Science 
of Vietnam for financial support to this work.

\begin{em}

\end{em}

\end{document}